\documentclass{webofc}
\usepackage[varg]{txfonts}   % Web of Conferences font
\usepackage{hyperref}
\usepackage{url}
\usepackage{caption}
\usepackage{subcaption}

\newcommand{\xt}{{\mathbf{x}_\perp}}

\hypersetup{colorlinks=true,citecolor=blue,urlcolor=blue,linkcolor=blue}
\begin{document}
\title{Perturbative high-energy evolution in the IP-Glasma initial state}
%
% subtitle is optionnal
%
%%%\subtitle{Do you have a subtitle?\\ If so, write it here}

\author{\firstname{Heikki} \lastname{Mäntysaari}\inst{1,2}\fnsep\thanks{\email{heikki.mantysaari@jyu.fi}} \and
        \firstname{Björn} \lastname{Schenke}\inst{3}\fnsep
        %\thanks{\email{Mail address for second
        %     author if necessary}} 
             \and
        \firstname{Chun} \lastname{Shen}\inst{4}\fnsep
        %\thanks{\email{Mail address for last
        %     author if necessary}}
        \and 
              \firstname{Wenbin} \lastname{Zhao}\inst{5,6}\fnsep
              %\thanks{\email{Mail address for last
             %author if necessary}}
        % etc.
}

\institute{Department of Physics, University of Jyv\"askyl\"a, P.O. Box 35, 40014 University of Jyv\"askyl\"a, Finland
\and
Helsinki Institute of Physics, P.O. Box 64, 00014 University of Helsinki, Finland 
\and
           Physics Department, Brookhaven National Laboratory, Upton, NY 11973, USA 
\and
           Department of Physics and Astronomy, Wayne State University, Detroit, Michigan 48201, USA
        \and 
        Physics Department, University of California, Berkeley, California 94720, USA
\and Nuclear Science Division, Lawrence Berkeley National Laboratory, Berkeley, California 94720, USA
          }

\abstract{We include the perturbative JIMWLK energy evolution into the IP-Glasma initial state description used to simulate the early-time dynamics in heavy ion collisions. By numerically solving the JIMWLK equation on an event-by-event basis, we obtain the energy (Bjorken-$x$) dependent structure of the colliding nuclei. Combining the initial state with hydrodynamic simulations, this enables us to predict how observables evolve when moving from RHIC to LHC energies.
}
\maketitle
\section{Introduction}
\label{intro}

Extracting the properties of the Quark Gluon Plasma (QGP), the deconfined phase of hot QCD matter, is a major objective of the heavy ion program at the Relativistic Heavy Ion Collider (RHIC) and the Large Hadron Collider (LHC). QGP properties can be inferred by comparing simulations with the measurements of bulk observables. % such as particle correlations. 
Hydrodynamical models have been very successful in describing such measurements from RHIC and LHC~\cite{Heinz:2024jwu}.

Simulations based on relativistic hydrodynamics need to be complemented by a description of the initial state. We focus on the impact parameter-dependent glasma (IP-Glasma) initial state model~\cite{Schenke:2012wb}. It has been succesfully used to describe RHIC and LHC heavy ion data~\cite{Schenke:2020mbo}, with the energy dependence obtained by  an energy-dependent saturation scale $Q_s$, which itself is parametrized and fitted to proton structure function data in the IP-Sat model~\cite{Kowalski:2003hm}.

In this proceeding we present results based on calculations shown in Ref.~\cite{Mantysaari:2025tcg}, where we move beyond this parametrized approach. Instead, the energy dependence is incorporated based on the perturbative JIMWLK equation~\cite{Mueller:2001uk}. This evolution requires non-perturbative input, describing the structure of the nucleus at the moderately small momentum fraction $x_0$, and then predicts the evolution towards smaller $x$. 

%A similar approach to predict the energy or rapidity dependence of the initial state geometry has been used e.g. in Refs.~\cite{Schenke:2016ksl,Schlichting:2014ipa,McDonald:2023qwc,Mantysaari:2023qsq,Mantysaari:2024qmt}.

\begin{figure}[htbp]
\centering
\begin{minipage}{0.48\textwidth}
  \centering
  \includegraphics[width=\linewidth]{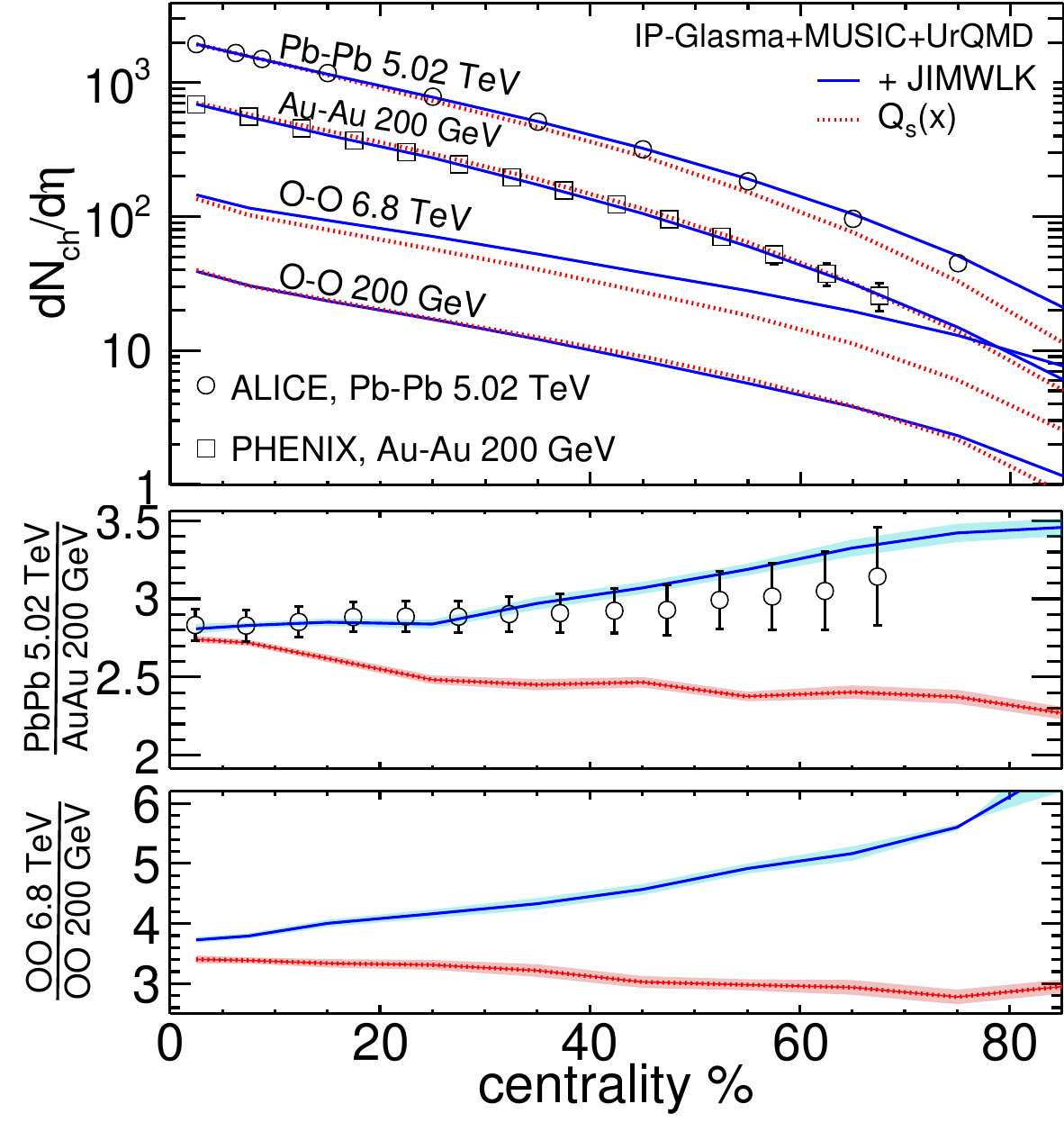}
  \caption{Charged hadron multiplicity distribution compared to ALICE and PHENIX data~\cite{ALICE:2015juo,PHENIX:2004vdg}. Figure from Ref.~\cite{Mantysaari:2025tcg}.}
  \label{fig:multiplicity}
\end{minipage}\hfill
\begin{minipage}{0.48\textwidth}
  \centering
  \includegraphics[width=\linewidth]{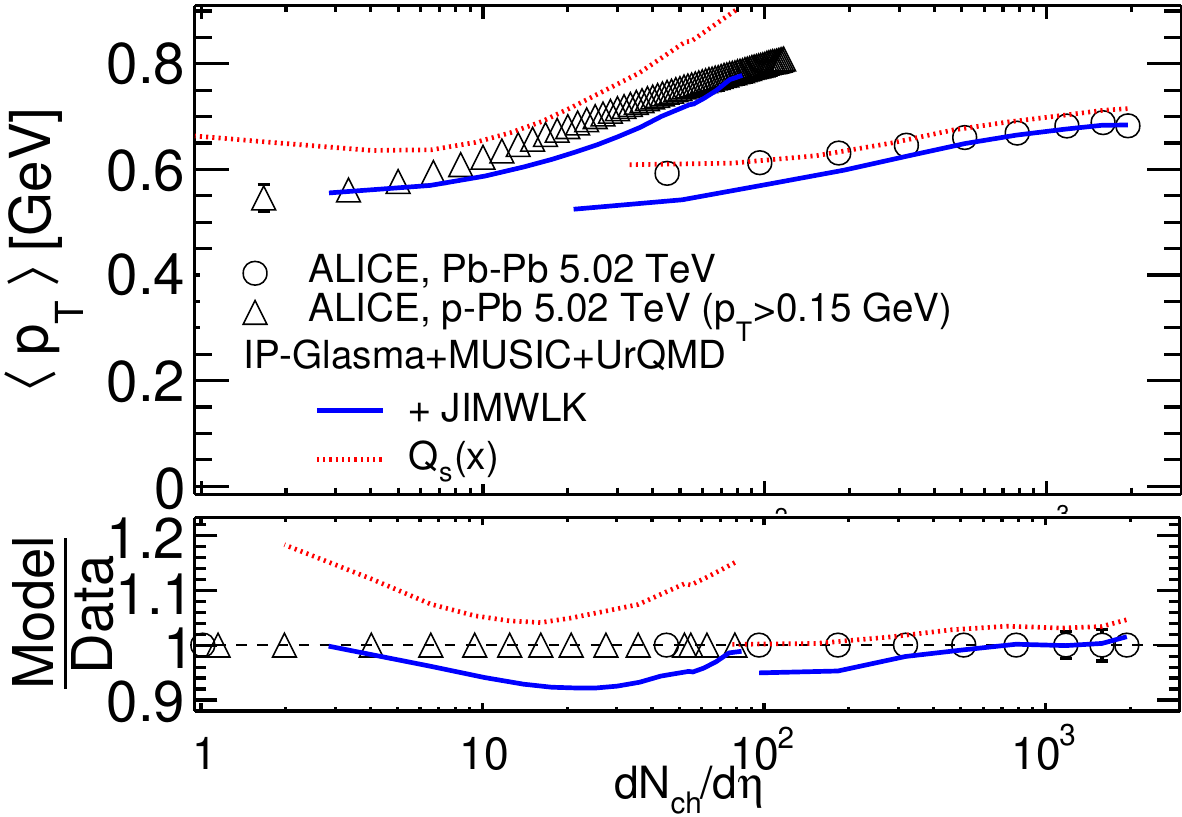}
  \caption{Average transverse momentum in proton-lead and lead-lead collisions at midrapidity compared to ALICE data~\cite{ALICE:2018hza}. Figure from Ref.~\cite{Mantysaari:2025tcg}.}
  \label{fig:meanpt}
\end{minipage}
%\caption{A figure with two panels.}
\label{fig:test}
\end{figure}

% \begin{figure}
% \centering
% \begin{subfigure}{.5\textwidth}
%   \centering
%   \includegraphics[width=\linewidth]{dNch_deta_PbPb_AuAu_OO.pdf}
%   \caption{\cite{ALICE:2015juo,PHENIX:2004vdg}.}
%   \label{fig:multiplicity}
% \end{subfigure}%
% \begin{subfigure}{.5\textwidth}
%   \centering
%   \includegraphics[width=\linewidth]{mean_pT.pdf}
%   \caption{The experimental data is taken from \cite{ALICE:2018hza}.}
%   \label{fig:meanpt}
% \end{subfigure}
% \caption{A figure with two subfigures}
% \label{fig:test}
% \end{figure}

\section{High-energy evolution for the initial state}

The IP-Glasma model is used to describe the proton and nuclear structure at the initial $x_0=0.01$. In this model, the degrees of freedom describing the highly energetic nuclei are Wilson lines $V(\xt)$, that describe the eikonal propagation of a quark through the color field of the nucleus at a given transverse coordinate $\xt$. These Wilson lines are then evolved to smaller $x$ by solving the JIMWLK equations.

The free parameters of the model control the overall saturation scale of the nucleon, the nucleon size, the nucleon substructure (implemented as three hot spots)~\cite{Mantysaari:2016ykx}, saturation scale fluctuations, coordinate space running coupling scale, and infrared regulators suppressing unphysical long-distance Coulomb tails. These parameters are constrained in Ref.~\cite{Mantysaari:2022sux} by comparing with the exclusive vector meson production data from HERA and from the LHC (a more recent analysis~\cite{Mantysaari:2025ltq} including uncertainty estimates was also presented at QM2025). 

When simulating heavy ion collisions, we solve the JIMWLK equation event-by-event down to $x=\langle p_T\rangle / \sqrt{s_\mathrm{NN}}$. Otherwise the setup is identical to the one used, for example, in~\cite{Schenke:2020mbo}. Early time evolution up to proper time $\tau_0=0.4$ fm/c is obtained by solving the  Yang-Mills equations implemented in IP-Glasma, after which the energy-momentum tensor $T^{\mu\nu}$ is computed and used to initialize the hydrodynamical phase, which is simulated using MUSIC~\cite{Schenke:2010nt}. The fluid cells are converted into hadrons at $e_\mathrm{sw} = 0.18$~GeV/fm$^3$, that are fed into the hadronic transport model (UrQMD) for further scatterings and decays~\cite{Bass:1998ca}. 

\section{Results}

Charged hadron multiplicity in Lead-Lead, Gold-Gold and Oxygen-Oxygen collisions are shown in Fig.~\ref{fig:multiplicity}. For comparison, we also use a setup labelled as ``$Q_s(x)$'', where the energy dependence only enters via the energy-dependent saturation scale $Q_s$. In this setup there is no geometry evolution, which in the case of JIMWLK evolution renders the nucleus smoother towards smaller $x$. At RHIC energies, both setups result in almost identical distributions, but significant differences are observed in the LHC energy range.

Fig.~\ref{fig:multiplicity} demonstrates that the center-of-mass energy dependence of the multiplicity distribution when moving from RHIC to LHC energy is well captured by the JIMWLK equation. We also show predictions for the oxygen-oxygen collisions, which is the only nucleus that will be measured at both RHIC and LHC energies. Due to the smaller system size, oxygen-oxygen collisions are more sensitive to the geometry evolution described by JIMWLK, and as such the differences between the two setups are enhanced in that case.

The average transverse momentum $\langle p_T\rangle$ as a function of midrapidity charged particle multiplicity is shown in Fig.~\ref{fig:meanpt}. Here results are shown both for Pb+Pb and p+Pb collisions, and again we compare the setups with and without JIMWLK evolution. As the JIMWLK evolution results in smoother nuclei, pressure gradients are smaller and consequently $\langle p_T\rangle$ is reduced. The available ALICE data~\cite{ALICE:2018hza} prefers the setup with JIMWLK evolution.

\begin{figure}
\centering
\begin{subfigure}{.5\textwidth}
  \centering
  \includegraphics[width=\linewidth]{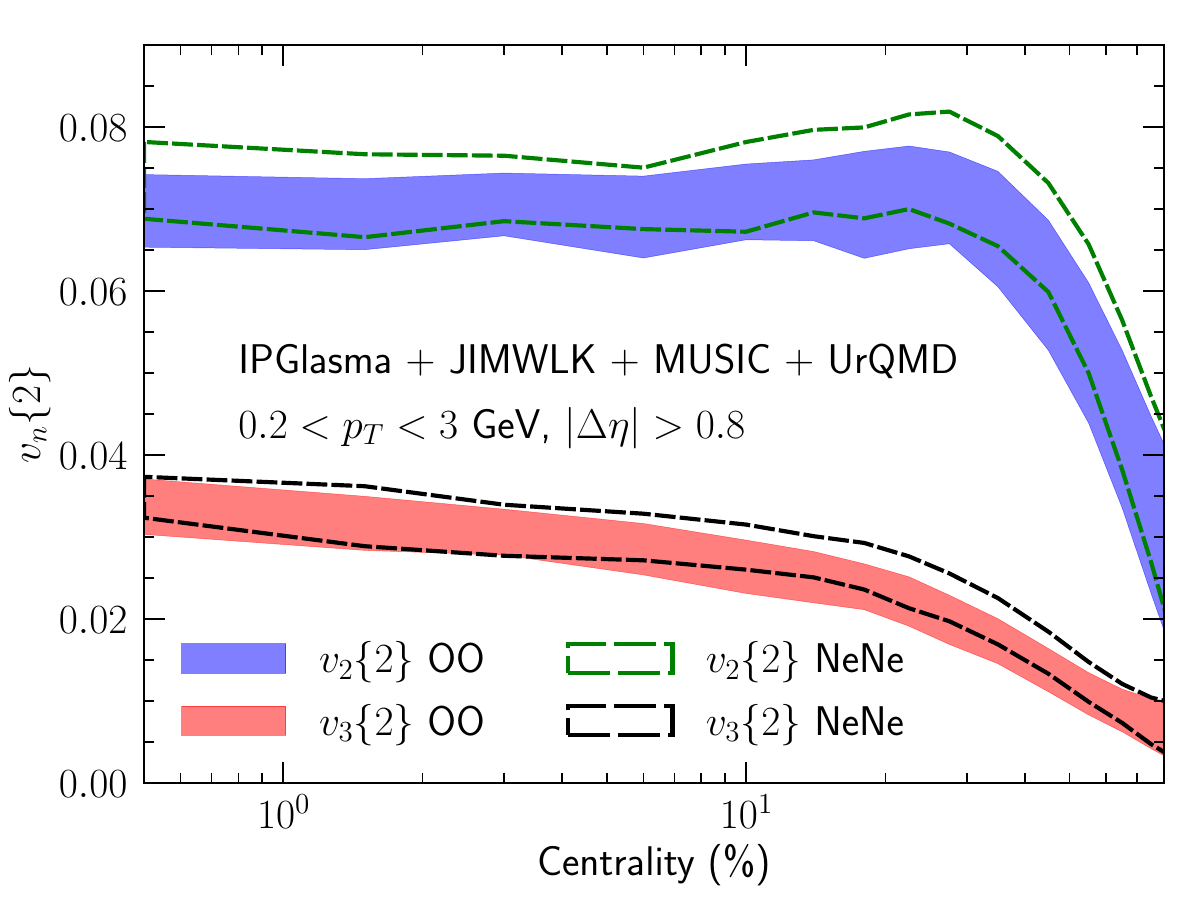}
  \caption{Centrality dependence of $v_2$ and $v_3$.}
  \label{fig:vn}
\end{subfigure}%
\begin{subfigure}{.5\textwidth}
  \centering
  \includegraphics[width=\linewidth]{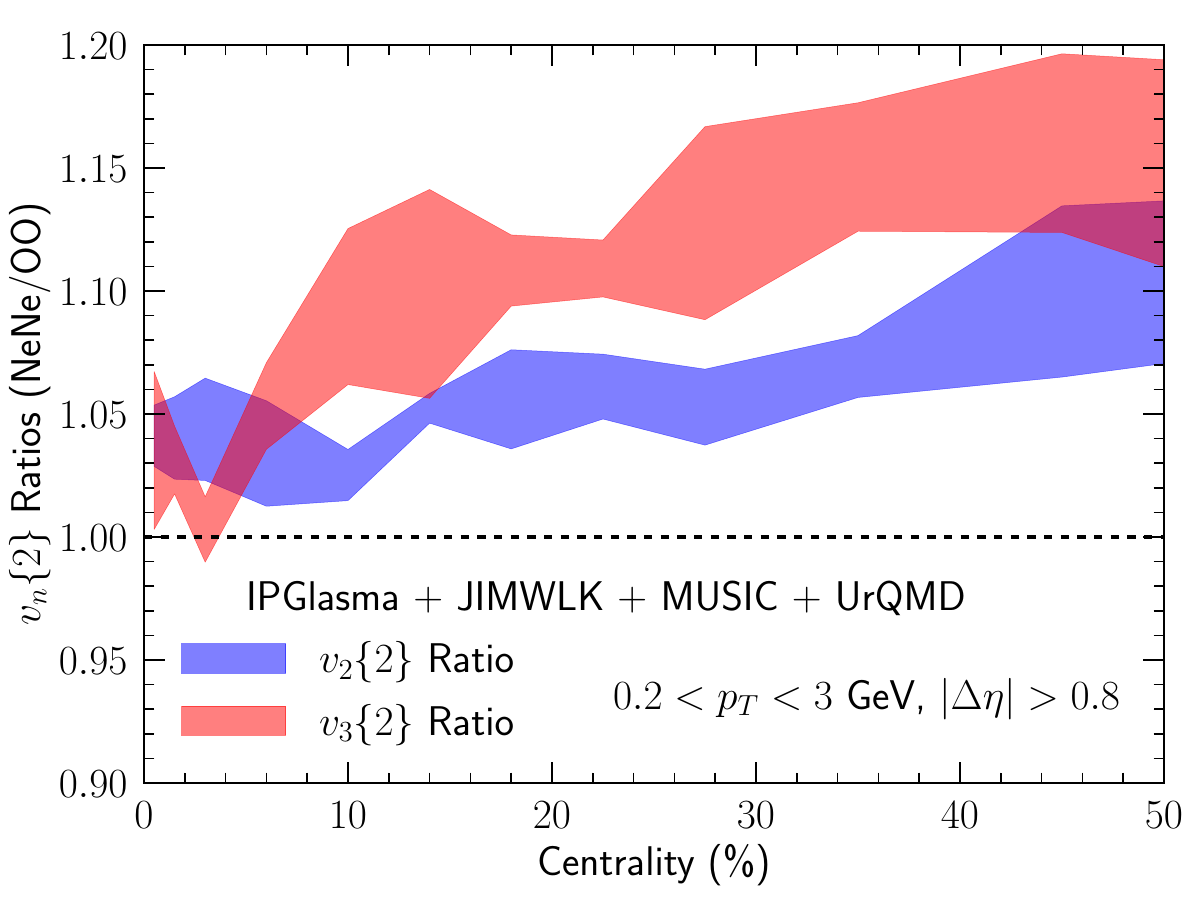}
  \caption{Ne+Ne/O+O ratio of $v_n\{2\}$ (n=2, 3).}
  \label{fig:vnratio}
\end{subfigure}
\caption{Elliptic and triangular flow in O+O and Ne+Ne collisions at 5.36 TeV. The uncertainty bands include systematic variation of the pre-equilibrium condition for initializing hydrodynamics, namely matching full $T^{\mu\nu}$, setting initial bulk viscous pressure $\Pi = 0$, and depositing only energy density at an early proper time. Those variations largely cancel between the two systems in the $v_n\{2\}$ ratios.}
\label{fig:vn}
\end{figure}

As an application of our setup, we calculate elliptic and triangular flow as a function of centrality in light ion collisions. We focus on oxygen-oxygen and neon-neon collisions at LHC energies $\sqrt{s_\mathrm{NN}}=5.36$ TeV. Our predictions are shown in Fig.~\ref{fig:vn}, where we also show the Ne+Ne/O+O ratio, for which many systematic uncertainties are expected to cancel.

\section{Conclusions}
We have included JIMWLK evolution into the widely used IP-Glasma initial state description. The developed publicly available~\cite{ipglasma_jimwlk_code} setup enables one to predict the center-of-mass energy dependence of the initial state description when simulating heavy ion collisions. 

We have demonstrated that even the simplest bulk observables, like multiplicity distribution and mean transverse momentum, are sensitive to the high-energy evolution in the initial state model. Consequently, an accurate and theoretically motivated energy evolution is necessary to precisely extract the fundamental properties of the QGP.

\bigskip
\noindent {\it{Acknowledgments.}}
{%\small 
B.P.S. and C.S. are supported by the U.S. Department of Energy, Office of Science, Office of Nuclear Physics, under DOE Contract No.~DE-SC0012704 and Award No.~DE-SC0021969, respectively.  C.S. acknowledges a DOE Office of Science Early Career Award. This material is based upon work supported by the U.S. Department of Energy, Office of Science, Office of Nuclear Physics, within the framework of the Saturated Glue (SURGE) Topical Theory Collaboration.
H.M. is supported by the Research Council of Finland, the Centre of Excellence in Quark Matter, and projects 338263 and 359902, and under the European Research Council (ERC-2023-101123801 GlueSatLight and ERC-2018-ADG-835105 YoctoLHC).
W.B.Z. is supported by the National Science Foundation (NSF) under grant number ACI-2004571 within the framework of the XSCAPE project of the JETSCAPE collaboration.
%This research was done using resources provided by the Open Science Grid (OSG)~\cite{Pordes:2007zzb, Sfiligoi:2009cct}, which is supported by the National Science Foundation award \#2030508.
The content of this article does not reflect the official opinion of the European Union and responsibility for the information expressed therein lies entirely with the authors.
}

%
% BibTeX or Biber users please use (the style is already called in the class, ensure that the "woc.bst" style is in your local directory)
 \bibliography{refs} % Replace "your_bib_file" with the actual name of your .bib file
%
% Non-BibTeX users please use
%
% \begin{thebibliography}{}
% %
% % and use \bibitem to create references.
% %
% \bibitem{RefJ}
% % Format for Journal Reference
% Journal Author, Article title. Journal \textbf{Volume}, page numbers (year). \url{https://doi.org/Article-DOI-number}
% % Format for books
% \bibitem{RefB}
% Book Author, \textit{Book title} (Publisher, place, year) page numbers
% % etc
% \end{thebibliography}

\end{document}